# Gyrokinetic projection of the divertor heat-flux width from present tokamaks to ITER

C.S. Chang[1], S. Ku[1], A. Loarte[2], V. Parail[3], F. Köchl[4], M. Romanelli[3], R. Maingi[1], J.-W. Ahn[5], T. Gray[5], J. Hughes[6], B. LaBombard[6], T. Leonard[7], M. Makowski[8], J. Terry[6]

[1]*Princeton Plasma Physics Laboratory, Princeton University, Princeton, NJ 08543-451, USA*
[2]*ITER Organization, Route de Vinon sur Verdon, 13067 St Paul Lez Durance, France*
[3]*Culham Centre for Fusion Energy, Culham Science Centre, Abingdon OX14 3DB, UK*
[4]*Atominstitut, Technische Universität Wien, Stadionallee 2, 1020 Vienna, Austria*
[5]*Oak Ridge National Laboratory, Oak Ridge, TN, USA*
[6]*Massachusetts Institute of Technology, Cambridge, MA, USA*
[7]*General Atomics, San Diego, CA, USA*
[8]*Lawrence Livermore National Laboratory, Livermore, CA, USA*

*Corresponding author*: cschang@pppl.gov

**Abstract:**

The XGC1 edge gyrokinetic code is used to study the width of the heat-flux to divertor plates in attached plasma condition. The flux-driven simulation is performed until an approximate power balance is achieved between the heat-flux across the steep pedestal pressure gradient and the heat-flux on the divertor plates. The simulation results compare well against the empirical scaling $\lambda_q \propto 1/B_P^\gamma$ obtained from present tokamak devices, where $\lambda_q$ is the divertor heat-flux width mapped to the outboard midplane, $\gamma=1.19$ as found by T. Eich *et al.* [Nucl. Fusion 53 (2013) 093031], and $B_P$ is the magnitude of the poloidal magnetic field at the outboard midplane separatrix surface. This empirical scaling predicts $\lambda_q \lesssim 1$mm when extrapolated to ITER, which would require operation with very high separatrix densities ($n_{sep}/n_{Greenwald} > 0.6$) [Kukushkin, A. et al., Jour. Nucl. Mat. **438** (2013) S203] in the Q=10 scenario to achieve semi-detached plasma operation and high radiative fractions for acceptable divertor power fluxes. Using the same simulation code and technique, however, the projected $\lambda_q$ for ITER's model plasma is 5.9 mm, which could be suggesting that operation in the ITER Q=10 scenario with acceptable divertor power loads may be obtained over a wider range of plasma separatrix densities and radiative fractions. The physics reason behind this difference is, according to the XGC1 results, that while the ion magnetic drift contribution to the divertor heat-flux width is wider in the present tokamaks, the turbulent electron contribution is wider in ITER. Study will continue to verify further this important projection. A high current C-Mod discharge is found to be in a mixed regime: While the heat-flux width by the ion neoclassical magnetic drift is still wider than the turbulent electron heat-flux width, the heat-flux magnitude is dominated by the narrower electron heat-flux.

## 1. Introduction

A challenge for ITER operation is the ability of the divertor to withstand the steady plasma exhaust heat that will be deposited on the divertor surface along a narrow toroidal strip. A simple, data-based regression from experimental measurements in present devices shows that the heat-flux width follows a scaling $1/B_P^\gamma$ where $B_P$ is the magnitude of the poloidal magnetic field on the outboard midplane separatrix surface and $\gamma \sim 1$. For ITER H-mode operation at $I_P$=15 MA with $q_{95} = 3$, this regression yields $\lesssim 1$mm for the heat-flux width $\lambda_q$ when mapped to the outboard midplane. For the range of $\lambda_q < 5$ mm in ITER, the peak divertor power fluxes in attached divertor conditions are beyond the design limits of the stationary heat loads of the ITER divertor target (10 MWm$^{-2}$), thus requiring the divertor operation in semi-detached conditions in which the plasma power is dissipated over a larger area by atomic radiation from hydrogenic-isotope atoms and impurities at the divertor. The operational range in which such semi-detached divertor operation can be achieved decreases with smaller $\lambda_q$, and is restricted to very high plasma separatrix densities and radiative fractions when $\lambda_q \sim 1$ mm ($n_{sep}/n_{GW} > 0.6$ for $\lambda_q \approx 1$mm) [1], where $n_{GW}$ is the plasma density in the core inside the pedestal top above which value the plasma tends to disrupt [2]. This raises concerns regarding their compatibility with the good H-mode energy confinement required to achieve Q=10 operation in ITER. In addition such a small $\lambda_q$ poses additional challenges for the control and sustainment of the semi-detached divertor conditions. The power fluxes during transient re-attachment exceed by a multiplication factor of several over the stationary heat flux design limits of the ITER divertor.

However, it is questionable if such an extrapolation from the present experimental data regression is valid as there may be differences in the fundamental heat-flux physics between ITER plasma and those in the present devices. Therefore, any extrapolation from present experiments to ITER may need to be on a more fundamental, kinetic physics basis, which is the purpose of the present XGC1 [3-6] studies. Firstly, the heat-flux width prediction from the XGC1 gyrokinetic model is to be validated against experimental data from the existing tokamak plasmas. Secondly, if the validation is satisfactory, the same XGC1 model can then be used to predict the heat-flux width in a model ITER edge plasma, with the caveat that the model ITER edge plasma may not be exactly what will be produced by ITER experiments. The model ITER edge plasma is an approximately power-balanced, local gyrokinetic equilibrium found in XGC1 after a manual profile adjustment from the initial model profile.

The heat flux to the divertor plates arises from plasma transport across and along the magnetic separatrix surface the near-scrape-off layer. Combined roles of all the dominant physics phenomena that exist in this relatively cold and non-Maxwellian plasma region needs to be included self-consistently to determine $\lambda_q$. The minimum requirements for such multi-physics phenomena can be listed as 1) the large amplitude, nonlinear intermittent, and mostly electrostatic turbulent structures called "blobs [7,8]," 2) the self-generated mean electric field that controls both the blobby turbulence and the neoclassical dynamics, including the quasi-neutral potential variation along the magnetic field lines, 3) the neoclassical physics driven by magnetic drifts together with parallel flows in the mean electric field, including the X-point orbit loss physics [9,10], and 4) the ionization and charge-exchange interactions with neutral particles. These multi-physics phenomena interact nonlinearly and self-organize together to produce the physics observables: the divertor heat-flux width in this case. Since the "blobby" fluctuations around the magnetic separatrix and in the scrape-off layer are known to be mostly electrostatic [8], we perform electrostatic simulations in the present study. The electrostatic gyrokinetic model, with the neglect of the electromagnetic effect, in the present study of the divertor heat-flux width must be validated against the existing experimental data. If the validation is not satisfactory to a reasonable degree, we must use a better model; i.e., electromagnetic gyrokinetic model. In order to reduce the chance of having an accidental agreement, a wide range of experimentally relevant situations need to be simulated. We note here that among the simulation data chosen in the present work, the C-Mod discharges showed quasi-coherent modes that have been identified as separatrix-spanning electron drift-waves with large electrostatic component ($\delta n/n \sim 30\%$, locally), but also with interchange and electromagnetic contributions ($\delta B/B \sim 10^{-4}$) [11].

The chosen model for this study has the essential multiscale edge physics as described above, but also has limiting assumptions. It would be ideal, but unnecessary, if the edge gyrokinetic simulation could be carried out to a global experimental time scale so that it could achieve a flux-driven transport equilibrium throughout the whole plasma volume all the way into the magnetic axis. The gyrokinetic simulations on the present-day computers can only be carried out to a turbulence and collision time scale (~ms in the core and ~0.1ms in the edge plasmas) due to error propagation and computational resources availability. This limitation is not a showstopper, though, since a flux-driven local simulation is a well-established practice as long as a proper boundary condition is used. In the present study, the local region of interest is from just inside the separatrix surface in the steep gradient region and the material wall: we need to establish a proper power-balance between the inside boundary and the material wall. The closer the initial plasma profile is to the gyrokinetic equilibrium, the sooner the saturation is achieved. We find that an experimental equilibrium is close to the gyrokinetic equilibrium, as it should be. If we started the simulation from an arbitrary plasma profile, the present multi-machine study would not have been possible within the available computational resources.

In the present study with XCG1, as we are interested in determining the lowest value of $\lambda_q$, we only consider plasmas in the attached divertor regime; i.e. we do not include strong radiative loss processes, avoiding the atomic-physics dominated detached plasma conditions in front of the divertor plates. We also try to minimize the charge exchange processes in the divertor chamber in order to identify the role of the warm ions from inside the separatrix in the divertor heat-flux physics. A detached plasma simulation may not allow such a manipulation in the neutral particle birth location and, also, could require a much longer simulation time than the time scale considered in the present study due to short neutral mean-free-path that leads to a slow diffusive neutral particle equilibration with the high density divertor plasma.

A general decription of the scrape-off-layer (SOL) plasma is given in the tutorial book by Stangeby [12]. The picture of parallel heat flow competing against radial turbulence transport in setting the divertor heat-flux width has been presented by multiple authors (see for example Refs. [13, 14]). These references suggest that the physics time scale in settling the divertor heat-flux width is on the order $L/V_{heat,\|}$ or the turbulence settling time, whichever is longer; where L is the field-line distance to divertor plates (L~ qR was assumed in those references with a typical edge safety factor value q), and $V_{heat,\|}$ is the heat flow speed along magnetic field line. In the attached divertor regime studied here, according to the conventional wisdom used in the above references, $V_{heat,\|}$ is somewhat smaller than the "flux-limit" speed $(m_i/m_e)^{1/2}C_s$. Here $m_i$ and $m_e$ are the ion and electron mass, respectively, and $C_s$ is the ion sound speed. Since $L/V_{heat,\|}$ is then only on the order $10^{-6}$ seconds in the attached regimes of three major US tokamaks, the settling time for the divertor heat flux width will be determined by the settling of the edge turbulence around the magnetic separatrix surface. This is true even if we consider $V_{heat,\|}$ to be the parallel tail ion speed, which would yield $L/V_{heat,\|} \sim 10^{-5}$s. Thus, the length of the simulations we perform here will need to be until both the edge turbulence intensity and the heat-flux width settle ($<10^{-4}$s, as will be seen subsequently), with an adequate power-balance locally between just inside the separatrix and the divertor plate location. We note here that this time scale is much shorter than the core turbulence equilibration time ($\sim 10^{-3}$s).

The paper is organized as follows: Section 2 provides a brief introduction to the XGC1 edge gyrokinetic code. Section 3 describes validation of the model against several representative experimental data on the divertor heat-flux width from three major US tokamaks, together with physics interpretations. Section 4 presents a prediction for the divertor heat-flux width in a model ITER plasma. Conclusions and discussion are given in Sec. 5.

## 2. The XGC1 edge gyrokinetic code

XGC1 is a full-f (=total-f), particle-in-cell based, 5-dimensional gyrokinetic code, specialized in simulating the flux-driven global plasma in diverted magnetic field geometry bounded by material wall [3-7]. It includes gyrokinetic ions; optional drift-kinetic, gyrokinetic, fluid, or adiabatic electrons; neutral particles with a recycling coefficient, and charge-exchange and ionization interactions with the plasma; and radiative power loss, external heat and momentum sources. In this study, the drift-kinetic electron option is used. The flux-driven simulation can be performed with a local power-balance in an annulus type simulation [3] or with a global power-balance for the whole device simulation [15].

The physics focus domain in the present study is between just inside the magnetic separatrix surface and the material wall. Instead of placing a simple core-edge boundary and initiating a radial heat flux there, as was done in some earlier XGC1 simulations [3], in this study we treat the whole core volume as the inside simulation boundary. This soft "free boundary" approach is taken here since it has been found that the edge turbulence solution gets distorted if we remove turbulence at an artificial hard core-edge boundary, owing to the nonlocal multiscale interaction effect [15]. We then let the heat-flux just inside the separatrix to be generated naturally from the turbulence and neoclassical physics, driven by the XGC1-adjusted experimental plasma profile, instead of placing an artificial heat flux there. An experimental level heat source is placed in the core side plasma.

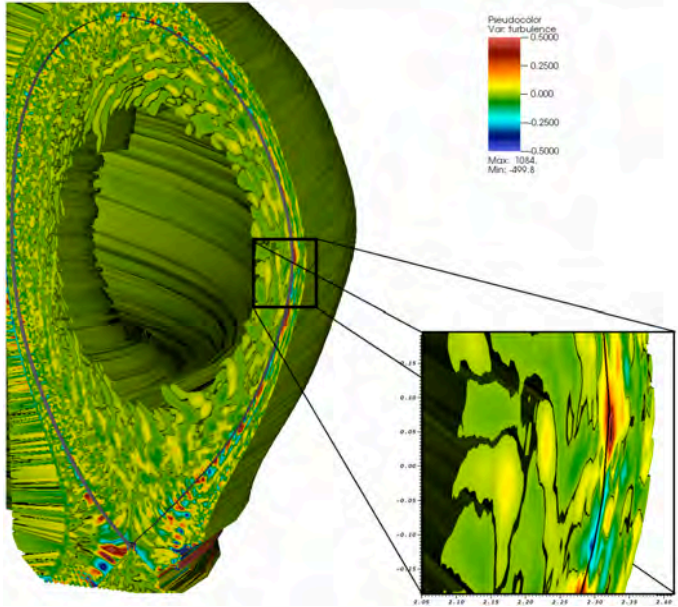

*Fig. 1. Typical electrostatic edge turbulence structure in a tokamak plasma. Blobby structure around the separatrix and in the scrape-off layer, and the ExB sheared streamer structures inward theric can be seen. Due to the openness of the inner boundary, turbulence in the core plasma is still developing even after the edge blobby turbulence has saturated. Visualization is by D. Pugmire of Oak Ridge National Laboratory in collaboration with the authors of this paper.*

Due to the strong turbulence activities together with strong sink (from large ExB shearing activities and parallel magnetic field-line connection to the conducting wall), the edge turbulence generation and saturation time scale is much shorter (<0.1ms or a couple ion toroidal transit time scale, see [3, 6]) than that in the core (~ a few ms, see [4]). Thus, by the time the turbulence in the vicinity of the separatrix is saturated, the turbulence in the core plasma may still be developing as can be seen from Fig. 1. Once the flux-driven full-f turbulence saturates in the edge region with an adequate power balance, its response to the core turbulence becomes stiff and weak even though the core-side turbulence is still developing, as learned from Refs. [3, 15] and will be shown here in a later section.

Figure 1 shows a representative example of the electrostatic edge turbulence solution, self-organized with neoclassical and neutral particle physics. These multi-physics dynamics are scale-inseparable. The thin black curve shows the magnetic separatrix surface with an X-point at the bottom. Near the separatrix surface ($\Psi_N \gtrsim 0.98$) and in the scrape-off layer (SOL), blobby turbulence structure [7,8] can be seen that is of non-streamer type (see the enlarged insert box around the outside midplane). Inside the blobby turbulence region ($\Psi_N < 0.98$), streamer type of large-scale turbulence structures driven by ion temperature gradient modes, trapped electron modes, and other drift-wave instabilities can be seen, sheared by the radial variation of ExB flow. The blobby edge turbulence has already saturated even before the central-core turbulence is generated. The approximately field-aligned structure of the turbulence ($k_\parallel \ll k_\perp$) can be seen at the inner and outer toroidal surfaces. The ion magnetic drift is oriented toward the magnetic X-point.

An extreme scale computing is required in the full-f edge plasma. Normally, an XGC1 edge simulation requires 3,000 – 10,000 particles per cell. For the ITER edge studies, we use 300 billion particles utilizing 90% of the 27 peta flop Titan [16] computer at ORNL for a few days. For the edge plasma studies in the present tokamak devices, the computational cost is smaller than that for ITER roughly in proportion to the problem size $(a/\rho_i)^2$, where a is the plasma minor radius and $\rho_i$ is the ion gyroradius. Thus, a DIII-D edge plasma for example study takes about 1 day of computing using 50% of Titan.

## 3. Validation against existing experimental data

### 3.A. General description of the simulation

Table 1 shows experimental discharges from three major US tokamaks that are used for validation of the XGC1 simulated divertor heat flux width. The cases were part of the 2016 US Department of Energy, Fusion Energy Science Theory/Simulation Milestone Target Research [17]. These discharges have been chosen to represent the wide range of the poloidal magnetic field $B_p$ on the outboard midplane separatrix surface as used for the all-tokamak regression study in Ref. [18, 19], with roughly equal spacing between the lowest and the highest $B_P$ values. All the chosen cases are in the single X-point geometry between or without ELMs with the magnetic drift directed toward the X-point.

| Shot | Time (ms) | $B_T$ (T) | $I_P$ (MA) | $B_{pol,MP}$ (T) |
|---|---|---|---|---|
| **NSTX 132368** | **360** | **0.4** | **0.7** | 0.20 |
| **DIII-D 144977** | **3103** | **2.1** | **1.0** | 0.30 |
| **DIII-D 144981** | **3175** | **2.1** | **1.5** | 0.42 |
| **C-Mod 1100223026** | **1091** | **5.4** | **0.5** | 0.50 |
| **C-Mod 1100223012** | **1149** | **5.4** | **0.8** | 0.67 |
| **C-Mod 1100223023** | **1236** | **5.4** | **0.9** | 0.81 |

*Table I. Experimental discharges from three US tokamaks that are used for validation of XGC results. $B_T$ is the toroidal magnetic field strength at magnetic axis, $I_P$ is the plasma current, and $B_P$ is the poloidal magnetic field strength at outboard midplane separatrix surface.*

Simulations performed in the present study are flux-driven, using the experimental level heat source input in the core plasma and the neutral particle recycling in the scrape-off plasma. The neutral atomic recycling coefficient is set to R=0.99 for a (approximate) conservation of the total plasma particle number inside the material limiter. R=0.99 is chosen to reflect some vacuum pumping effect in the divertor area. R=1 will not make a difference in the time scale of edge turbulence and neoclassical dynamics considered here. The neutral atomic recycling source is poloidally localized to the divertor area in the simulation.

In the present study of attached plasmas, the atomic recycling source is not placed right on the divertor plates. It is placed at the divertor entrance (horizontal X-point surface) in order to avoid concealing of plasma physics information by the strong neutral-particle atomic physics in the divertor chamber, while still preserving the total number of plasma particles. Generation of the neutral atoms near the divertor plates would raise the atomic neutral density in the divertor chamber (to a realistic level, though), and the resulting high rate of local charge exchange process would not allow us to study the physics of upstream ion contribution to the divertor heat-flux footprint. Thus, the ion heat-flux footprint that is evaluated from the present simulations does not fully include the local charge exchange effect right in front of the divertor plates. Because of the low rate of electron impact ionization at low electron temperature, the upstream electron contribution to divertor heat-flux width would be less affected by this manipulation of the atomic neutral birth distance from the divertor plates. This choice of neutral recycling location is not expected to influence sensitively the $\lambda_q$ evaluation in the present attached plasma studies for the reason explained in the next paragraph and in Sec. 3.C.

Moreover, the divertor entrance has been chosen here to be the horizontal surface crossing the X-point for consistency with the definition of the "power spreading parameter S" used by Eich et al. [18, 19], as to be summarized in Sec. 3.C of this paper. In Refs. [18, 19], it was assumed that an in-out symmetric diffusive and a dissipative spreading across the magnetic field lines, in addition to parallel and radial transport processes in the scrape-off layer, exist between the X-point and the divertor plate that is dependent on local divertor plasma properties (including the neutral density) and geometry. The parameter "S" is evaluated using the inward spread of the heat flux into the private flux region, and used to estimate the outward spread. Reference [20] developed this concept further and identified a relation between $\lambda_q$ and the actual heat-flux footprint on the divertor plates. The procedure in [18, 19] is taken in such a way that the local divertor plasma parameters and geometry do not influence the quantification of $\lambda_q$ by a convolution process. Thus, the artificial alteration of the local divertor plasma parameters from the de-emphasis of the neutral particle density is not expected to influence $\lambda_q$ much, even though it may alter the actual heat-flux footprint shape measured on the divertor plates.

We have performed a simplified sensitivity study of the divertor heat-flux width on the neutral particle density within our model. The neutral recycling has been switched on and off during a simulation. It is found that the full neutral particle density case can raise the divertor heat-flux width by up to 10% over the zero neutral particle density case in our attached plasma.

To save the computing resources, simulations in the present study are performed only until an approximate edge turbulence saturation just inside the magnetic separatrix surface and an adequate power-balance (among the core heating, power-flux just inside the separatrix and divertor plates) are achieved, which is much longer than the parallel thermal speed discussed at the end of Sec. 1; and also until the divertor heat-flux width is saturated. However, the simulation is not prolonged until the core turbulence is saturated due to the slowness of the core turbulence time scale compared to the edge turbulence time scale (see Fig. 1 and the associated discussions). Figure 2 shows a typical time behavior of the average turbulence intensity $<(\delta n/n)^2>$ at $\psi_N=0.989$ and the divertor heat-flux width $\lambda_q$, using the C-Mod 0.9MA shot #110030223023. Toward the end of the simulation, the average density fluctuation $<(\delta n/n)^2>^{1/2}$ settles to ~3% and $\lambda_q$ to ≈1mm ±10%. The blobby nonlinear turbulence starts at $\psi_N$≈0.98 and spreads radially outward while increasing its amplitude [7]: In the scrape-off layer $<(\delta n/n)^2>^{1/2}$ usually reaches well above 10%.

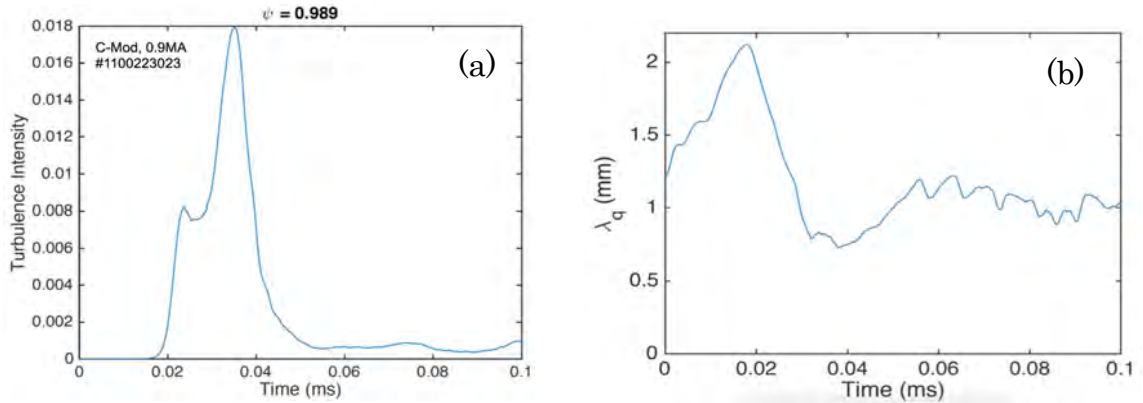

*Fig. 2. A typical time behavior of (a) average turbulence intensity $<(\delta n/n)^2>$ and (b) Eich's $\lambda_q$ in mm (see Sec. 3.C and Refs. [18,19] for the definition of $\lambda_q$). After the transient behavior is over at t~0.05ms, both turbulence and $\lambda_q$ saturate to quasi-steady level. The C-Mod shot #110030223023 with $I_p=0.9MA$ is used for this example.*

The heat flux across the steep pedestal gradient region at $\psi_N$=0.989 shows an approximate saturation around 1.2MW (Fig. 3), which is roughly the same net power (heating minus radiation loss) injected inside the surface $\psi_N$=0.989 and roughly the same power load on the divertor plates. Out of the 1.2MW heat flux, about 3/4 is from ions. The ion thermal transport is greater than the electron thermal transport due to the neoclassical transport. Thus, this flux-driven simulation has arrived at an approximate power balance in the physics domain of interest. Here, we call the power-balance in this simulation "approximate" since all the physical values in Figs. 2 and 3 fluctuate in time due to the turbulent nature of the system, and also their values at t=0.1ms can only be regarded as approximately saturated due to a bit premature ending of the valid simulation period. The simulation continued somewhat longer after that, showing longer saturation of these quantities, but with contamination after that by an unknown numerical instability localized to the magnetic separatrix.

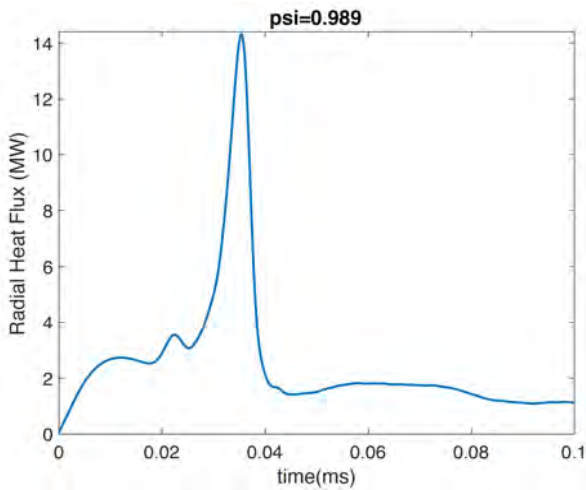

Fig. 3. Time behavior of the summed ion and electron heat flux through the $\psi_N$ =0.989 surface.

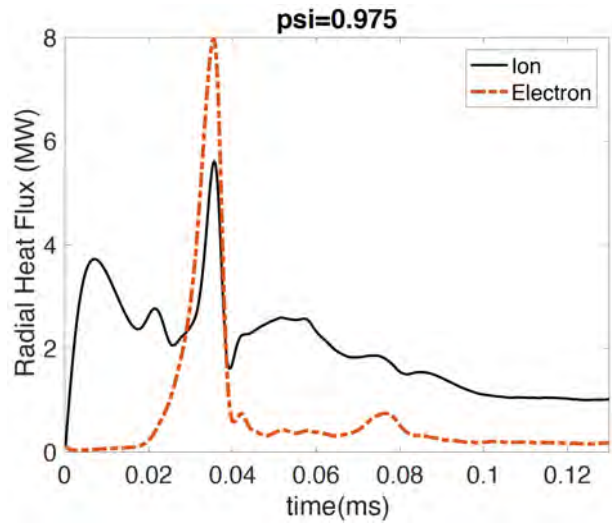

Fig. 4. Time behavior of the individual ion (solid line) and electron (dotted) heat fluxes through the $\psi_N$ =0.975 surface.

For the sake of completeness and being informative, we also plot in Fig. 4 the time behavior of the heat fluxes passing through an inner surface $\psi_N$=0.975 deeper into the pedestal. The indivicual ion and electron heat-fluxes are shown in Fig. 4. As alluded earlier, the turbulent heat-flux saturation begins to occur at a bit later time (≈0.1ms) compared to that on the $\psi_N$=0.989 surface. As explained earlier, the full-f flux-saturated solution begins at the outer radius and propagates inward. This delay in the turbulence saturation becomes greater as the observation surface moves inward, reaching to ≳1ms in the core plasma inside the pedestal top. Again, the difference between the ion heat-flux and the electron heat-flux is the neoclassical transport. Neoclassical radial electron transport is small.

The initial input plasma profiles from approximate experimental data and the approximately flux-balanced plasma profiles determined by XGC1 are plotted together in Fig. 5 for (a) electron density, (b) electron temperature, and (c) for ion temperature. The shift of the density pedestal towards the scrape-off layer region is due to shallow neutral ionization in this high density C-Mod pedestal (DIII-D and NSTX simulations do not show this behavior). A small ion temperature peak in the scrape-off region, reproduced by XGC1, is found to be due to the non-Maxwellian tail population. The pedestal plasma profiles stay close to the initial approximate experimental data. However, it is unknown how well the flux-balanced gyrokinetic plasma solutions (dotted lines) compare with the error limits from the experimental plasma profiles since the experimental measurement of the sharp edge pedestal profiles (with the width ≲ 1mm in this C-Mod case) is not possible due to the spatial resolution and other issues.

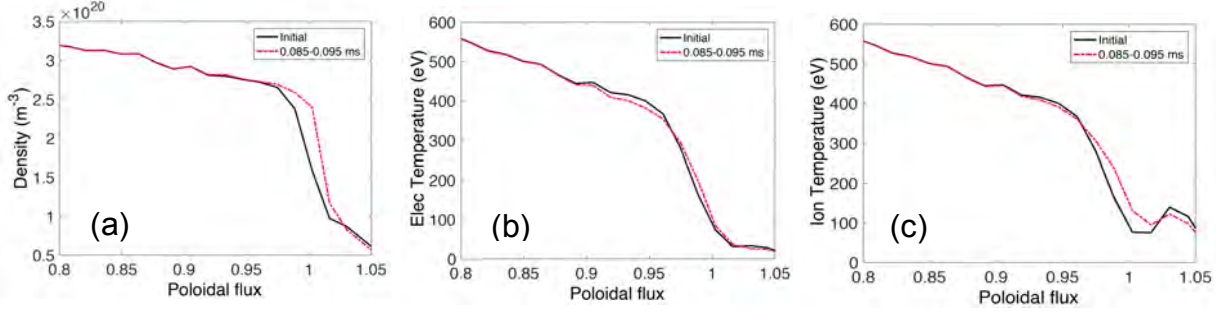

*Fig. 5. Initial (solid line) and the XGC1-evolved flux-balanced(dotted line) radial distribution of (a) electron density, (b) electron temperature, and (c) ion temperature for the edge of the C-Mod 0.9MA case. A small peak of ion temperature in the scrape-off region is from the non-Maxwellian tail population.*

### 3.B. Findings from the simulation of the present US tokamaks

Figure 6(a) depicts the ion and electron contributions to divertor footprint profile measured at the last grid points in front of the divertor plates, using another typical tokamak edge plasma (the DIII-D 1.5MA shot #144981). The parallel heat flux density is plotted as function of the outboard midplane distance $\Delta R_{mid}$ from the separatrix surface, with the grid points on the divertor plates mapped to the outboard midplane using the $\Psi_N$ contours. Negative distance means into the private flux region, but mapped to the corresponding $\Psi_N$ values inside the separatrix at the outboard midplane. Mapping by average flux expansion ratio, as done in some experiments, yields a similar result. Since the Debye sheath is treated as a subgrid phenomenon, using a logical sheath argument [21], effect of the Debye sheath is not included in the footprint analysis. The main effect of the Debye sheath would be the transfer of some electron kinetic energy to the ion kinetic energy by the amount of Debye sheath potential energy and will not affect the total ion and electron energy.

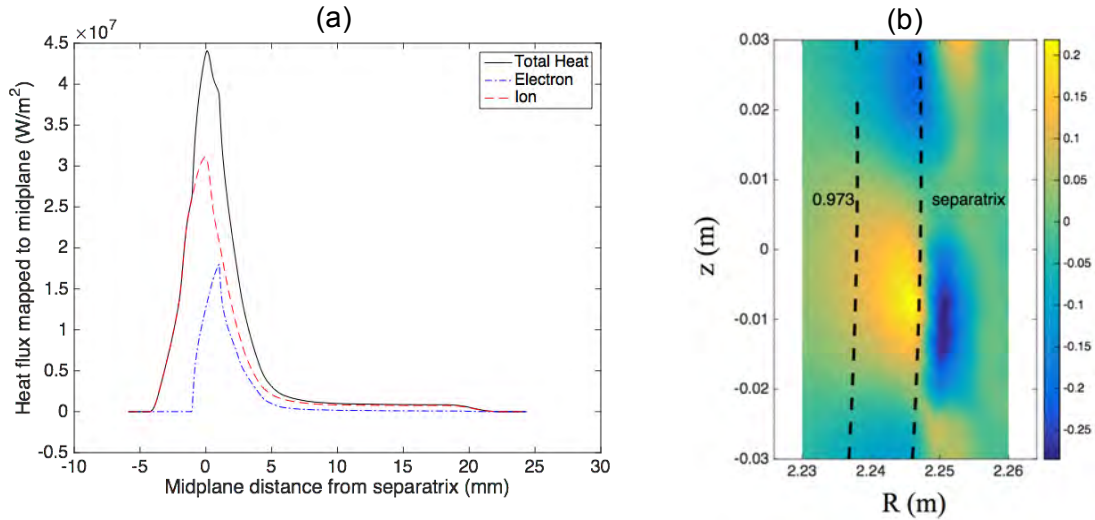

*Fig. 6. (a) Ion and electron heat flux footprints at the last grid point in front of divertor plates for the DIII-D 1.5MA shot #144981, are plotted in terms of the midpane distance from separatrix when the last grid point positions at divertor plates are mapped to the outboard midplane using the equal $\Psi_N$ values, and (b) $\delta n/n$ from intermittent turbulence across the separatrix surface around the outboard midplane is shown for comparison with the electron contribution pattern to the heat-flux footprint.*

It can be seen from this DIII-D case that the ion contribution to the divertor heat-flux, which contains both the neoclassical and some blobby turbulence effects, has a wider footprint and a greater magnitude than the electron footprint does. The electron contribution is predominantly from the blobby turbulence spread effect since the neoclassical magnetic drift effect is negligibly small. Figure 3(b) shows a snapshot picture of the blobby $\delta n/n$ around the outboard midplane for comparison with the electron heat-flux footprint shape. Z is the vertical distance and R is the horizontal major radius distance. It can be seen that the radial spread pattern of the intense amplitude part of blobs around the separatrix roughly agrees with the electron footprint spread pattern. All the DIII-D and NSTX cases obey the findings given in this and next paragraphs.

Due to the automatic quasi-neutrality of a gyrokinetic simulation, the particle flux (or electrical current) width is the same between electrons and ions (see Fig. 7). The wider heat-flux width of the ions than the electron heat-flux width means that a more significant portion of the ion heat-flux comes from the warm tail ions. Warm ions can travel from inside the sepratrix to the divertor plates via magnetic drift motions across the separatrix surface. Even though the background ions will be collisional in the scrape-off layer, the tail ions from the pedestal can be weakly collisional and their mean free path can be non-negligible compared to the device size. In this gyrokinetic physics picture, the warm ion flow to the divertor is not along the magnetic field lines from the upstream scrape-off layer plasma, but along the path combined with the ExB and magnetic drift motions that make the ion orbit migrate between magnetic field lines. The combined particle motions in XGC1 shows that the time for the ions to travel from the separatrix surface to the divertor plates is on the order $R/v_i$, where $v_i$ is the ion particle speed, not $qR/v_i$ as conjectured in the previous reduced models introduced in Sec. 1. The combined ion motions make the individual ions not to feel the effect of the small local poloidal magnetic field effect ($B_P/B \rightarrow 0$) in the vicinity of the X-point. In the three US tokamaks the time scale for the particle convection to the divertor plates is $R/v_i \sim 0.01$ms, which is shorter than the present simulation time scale. As the ions execute the combined motions, the quasineutrality and the ambipolar flow to the divertor plates are satisfied in corporation with the turbulent radial electron spread, to be discussed shortly below.

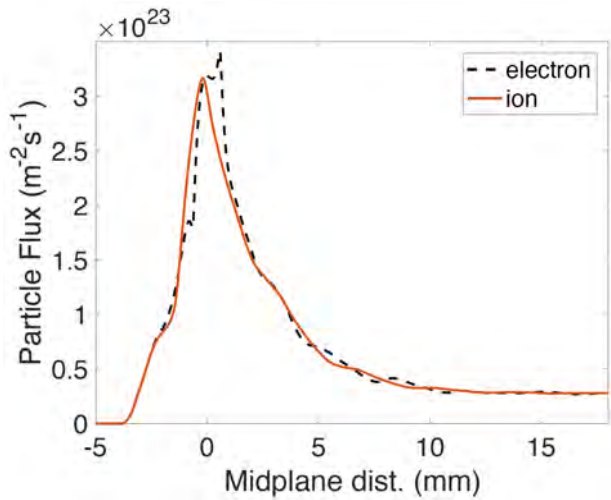

*Fig. 7. Gyrocenter particle-flux balance between ions and electrons on divertor plates, mapped back to the outboard midplane in the 1.5MA DIII-D case. The difference is from the ion polarization density. Since the gyrokinetic Poisson equation is built on charge neutrality, the physical particle fluxes are ambipolar.*

The other DIII-D cases and the NSTX case studied here lead to the same conclusion as described so far in this section. The C-Mod discharges, which have lower pedestal ion temperature and, thus, higher collisionality, yield somewhat different results, as to be shown later in this section. We note here again that if we located the neutral particle source near the divertor plate, we would not have been able to identify the origin of the heat-flux width so clearly between electrons and ions due to the strong local charge exchange effect that would conceal the upstream ion contribution to the divertor heat-flux width.

At this point, some heuristic discussions on the turbulent transport across the magnetic separatrix surface and the scrape-off layer (SOL) can be helpful before the presentation of the C-Mod and ITER results. The turbulent transport mechanisms in the scrape-off layer can be significantly different from those in the core plasma. Across the nested flux surfaces in a core plasma, the turbulent δExB convective cell motions of plasma particles may not lead to non-ambipoar radial transport since ExB motion is common to both electrons and ions. Moreover, a perfectly coherent δExB convective cell motion will not lead to time-averaged radial transport in a core plasma. However, in a boundary plasma in which particle orbits intersect with the material wall, different orbit dynamics between electrons and ions can make them feel different de-correlation of the δExB motion – thus lead to different radial transport for each species. The much faster parallel thermal motions of electrons can more easily break the coherence in the δExB convective motions and lead to electron radial random walk that is more sensitive to the quasi-coherent blobby turbulence than the ions are. As a result, the blobby turbulence effect is expected to be reasonably represented by the electron heat-flux footprint, while it is less important for the ion heat-flux footprint. This mechanism will also break the automatically ambipolar ExB convective transport across the magnetic separatrix and in the scrape-off layer, and can modify the mean neoclassical electric field profile to recover the ambipolarity when the ions experience the neoclassical magnetic and ExB drift from inside the separatrix towards the divertor plates. The random-walk transport physics by the δExB convective cell motions presented in this paragraph is not well-known to the general readers, and is described here to add to the usual argument that the radial plasma transport by blobs in the scrape-off layer is convectively carried by radial blob motion.

In a plasma with a strongly sheared ExB flow, turbulence tends to be compressed to the ExB shearing layer width through the shear de-correlation (see Refs. [3, 22] and the references quoted therein). In a diverted H-mode edge layer, the dominant mechanism for generation of a strongly sheared ExB flow is the ion orbit loss that is caused by the strong magnetic shear near the X-point [9,10,23,3] (the safety factor q diverges logarithmically toward the separatrix), with some modification by the non-ambipolar turbulence transport across the separatrix surface as described in the previous paragraph. Since the turbulent blobs are found to be generated just inside the magnetic separatrix surface ($\Psi_N \approx 0.98$) in the XGC1 simulations and move out radially across the scrape-off layer, their radial size may roughy be estimated by assessing the magnetic shear-layer width where they are formed and spend their young lives (just inside the separatrix).

From the X-point orbit loss physics [9], the radial ExB shear layer width $\Delta r$ just inside the separatrix surface ($0.98 \lesssim \Psi_N <1$) can be estimated from the relation "vertical drift time across the shear layer width $\Delta r_x$ above the X-point ~ poloidal motion time across the distance a" or $\Delta r_x/V_B \sim a/(v_{i\|m}B_{\theta x}/B)$ near the X-point, where $v_{i\|m}$ is the minimal parallel speed of the orbit-lost ions and "a" is the horizontal minor radius of the magnetic separatrix. The poloidal magnetic flux conservation yields $\Delta r_x B_{\theta x}=\Delta r B_\theta$, where $\Delta r$ and $B_\theta$ are their respective midplane values. Defining $v_{i\perp}$ to be the minimal perpendicular speed of the orbit-lost ions, this relationship trivially leads to

$$\Delta r \sim q(\Delta r)\, \rho_i\, (v_{i\perp}/v_\|), \qquad (2)$$

just inside the magnetic surface, where $q(\Delta r)$ is is the magnetic safety factor at the distance $\Delta r$ from the magnetic separatrix, $\rho_i$ is the gyroradius of the orbit lost ions, and the value of $(v_{i\perp}/v_\|)$ depends on the magnetic geometry and normally a number a few to several. Equation (2) says that the ExB shearing layer width is not only dependent on the ion gyroradius, but also on the device size through the radial variation of the safety factor q just inside the magnetic separtrix surface where the blobs are born. If the device size is large, q varies more slowly in the minor radius r. The local magnetic twist factor $B/B_{\theta x}$ diverges logarithmically toward the X-point to dominate the behavior of the flux-function safety factor $q(\Delta r)$ in the near vicinity of the magnetic separatrix. Let us use $q \sim -log[\Delta r/\alpha a]$ as a simple model for the singular q-behavior close to the magnetic separatrix surface along the outboard midplane. Here the parameter $\alpha$ is a small number and makes $q=q_{95}$ at $\Psi_N=0.95$ outside of this singular shear layer. Dependence of $\alpha$ on magnetic equilibrium shape will be neglected in this simple argument. The actual value of $\alpha$ is not needed here. Eq. (2) then becomes

$$\Delta r \sim -\rho_i\, log[\Delta r/\alpha a]. \qquad (3)$$

Since $log[\Delta r/\alpha a]$ is not an algebraic function and difficult to get a scaling idea, let us use a rough algebraic approximation at some finite distance from the separatrix, $log[\Delta r/\alpha a] \sim (\alpha a/\Delta r)^k$. (This approximation will be worst at an infinitesimal distance from the separatrix surface.) From the mathematical relation $1+|log[x]| \leq 1/x$ for $x<1$, we can see that the maxmum value of k is 1 (k actually would be a bit smaller than 1). By using k=1 in Eq. (3), we obtain the following approximate scaling for the ExB shear layer width near the magnetic separatrix, which limits the radial blob width through the ExB shear decorrelation mechanism,

$$\Delta r \sim (\alpha a\, \rho_i)^{1/(1+k)} \sim (\alpha a\, \rho_i)^{1/2}. \qquad (4)$$

This argument shows that the radial "blob" width could be approximately proportional to $(a\rho_i)^{1/(1+k)} \sim (a\rho_i)^{1/2}$ and has a $(a)^{1/2}$ size-scaling, approximately. In the present US tokamaks, $(a)^{1/2}$ does not lead to a significant machine to machine variation. However, between C-Mod and ITER, the difference in $(a)^{1/2}$ can be significant.

The argument presented here is only for heuristic purpose, and should not be considered as a rigorous proof. Note the similarity between Eq. (4) and the meso-scale turbulence spreading distance $\sim (L_P \rho_i)^{1/2}$ that leads to Bohm-type turbulent transport [24], if we substitute the magnetic shear length $\sim \alpha a$ by the plasma pressure gradient length $L_P$.

A limiter discharge will also generate an orbit-loss-driven ExB shearing layer just inside the last closed magnetic surface. In this case, the ExB shearing scale length is about a banana width of the tail ions. A similar equation to Eq. (2) can be obtained, with an algebraic expression for the $q(\Delta r)$ profile. Further discussion on the limiter plasma is not within the scope of this study.

With this heuristic argument on some of the edge turbulence properties, we now go back to the numerical observation of the divertor heat-flux footprint for C-Mod. Ion and electron contributions to the divertor heat-flux footprint for the high current C-Mod discharge case ($I_p$=0.9MA, #1100223023) are plotted in Fig. 4. This case is shown here because it has the highest $B_P$ value among the regression data used by Eich et al. [18,19]. According to the existing argument that $k_\perp \rho_i$ is around 0.1-0.2 for intermittent blob envelopes in the present devices and according to Eq. (4), this case has the smallest blob size. It also yields the smallest neoclassical magnetic drift effect.

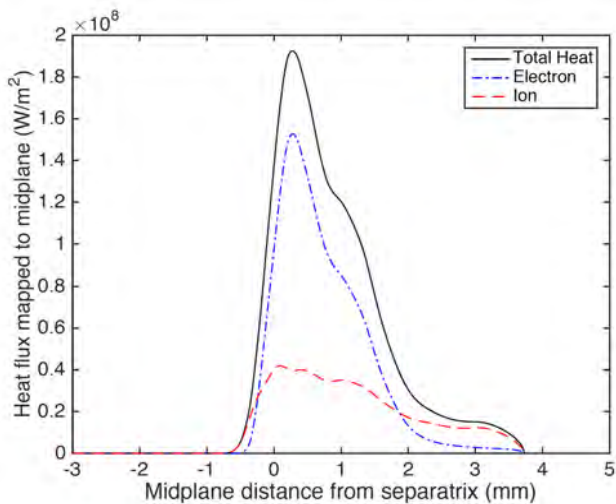

*Fig. 8. Divertor heat-flux footprint by electrons and ions in the high current C-Mod plasma discharge #1100223023 ($I_p$=0.9MA).*

It can be seen from Fig. 4 that the ion heat-flux footprint, which is mostly driven by the neoclassical magnetic drift of warmer ions, is still wider than that by electrons. However, the heat-flux magnitude is dominated by the electron contribution, which is driven by the blobby turbulence. Thus, the high current C-Mod case is in a somewhat different regime from the DIII-D and NSTX cases in that the turbulent electron thermal footprint width is still smaller than the ion magnetic drift width, but determines the total heat-flux width due to its dominance in magnitude. The low heat-flux magnitude by ions can easily be understood from the low pedestal $T_i$ in C-Mod and, thus, the warm tail ions have much less power. Moreover, higher collisionality during their travel from the pedestal to the divertor chamber can thermalize the tail ions more easily.

We note here that the decrease of the ion orbit width with toroidal current, the heuristic device size scaling in Eq. (4), and the electron heat-flux width following the blob size all together imply that $\lambda_q$ may decrease further at higher plasma current in C-Mod. The higher current C-Mod case is left for a future study.

### 3.C. Definition of the heat-flux width

Even for the same upstream plasma, the divertor heat-flux width can be different for different divertor plasma conditions and geometry. A unified definition is needed that is not dependent upon the divertor plasma condition and geometry. For that reason, we choose to follow the heat-flux width definition by Eich et al. [18], quoted here with the reference and equation numbers properly adjusted:

"*Assuming a purely exponential radial decay (characterized by $\lambda_q$) of the parallel energy transport, the inter-ELM outboard midplane SOL parallel heat flux profile can be written as $q(r) = q_\| Exp[-r/\lambda_q]$, where $r=R-R_{sep}$, $R_{sep}$ being the major radius of the separatrix at the outer midplane. We further assume that $\lambda_q$ is dependent only on the upstream outer midplane SOL parameters and the magnetic connection length along field lines to the outer target, $L_c$. Heat transport into the private flux region is included by describing the observed power spreading (diffusion/dissipation) along the divertor leg between the X-point and the target as a Gaussian spreading of a point heat source; this can simply be taken into account by convoluting q(r) with a Gaussian function of width S [25], which we refer to as the power spreading parameter and which is assumed to be dependent on local divertor plasma parameters and geometry. The result of this convolution is the following expression for the outer target profile [19]:*

$q(s*) = (q_0/2) \exp[(s/2\lambda_q)^2 − s*/(\lambda_q f_x)] \, erfc[S/(2\lambda_q) − s*/(S f_x)] + q_{BG}$, and (5)
$s* = s-s_0 = (R_{sep} − R) f_x$.

*The other quantities used in equation (5) are the background heat flux, $q_{BG}$, the effective flux expansion, $f_x$, on the target following the definition in [20], and the peak heat flux at the divertor entrance $q_\| = q_0/ sin(\theta_\perp)$ with $\theta_\perp$ the field line angle on the divertor target.*"

Reference [18] also gives an analytic regression formula for $\lambda_q$, obtained from data from all the major tokamak devices including the tight aspect ratio machines:

$\lambda_q^{(14)} = 0.63 \, B_{pol}^{-1.19}$ mm, (6)

with the square correlation being 0.86. The superscript "(14)" represents the regression #14 in Table 3 of Ref. [18]. $\lambda_q$ results from the XGC1 simulation will be compared against this regression formula. Figure 3 in Ref. [18] then shows the experimentally obtained power fall-off length ($\lambda_q$) versus poloidal magnetic field at the outer midplane separatrix, together with the regression error limits, for the all-machine experimental data they sampled.

One comment to be made here is that the interpretation of the "power spreading parameter" S in the simulation (and in the real situation) can be different from the ideal definition used in Ref. [18]. In Ref. [18], it is assumed that a purely diffusive spreading from the magnetic X-point into the private flux region determines S, and that the same spreading also occurs radially outward. In the simulation (and in the real situation), however, the inward spreading into the private flux region has a convective neoclassical magnetic drift component that is not symmetric with the radially outward drift due to the angle of the outer divertor separatrix leg relative to the magnetic drift direction.

Notice here that in this definition of $\lambda_q$, the divertor plasma condition and the geometry below the X-point is removed by the introduction of the parameter S. Making connection back to Section 3.A, the intentional de-emphasis of the neutral particle atomic physics in the divertor chamber below the X-point is thus expected to have a insignificant impact on the determination of $\lambda_q$, while allowing for the kinetic study of the upstream ion and electron effect on the divertor heat-flux footprint.

### 3.D. Comparison with experimental footprint

In this section, XGC1 predictions are compared against the representative experimental data. Figure 9 shows an example heat-flux footprint using the C-Mod 0.5MA discharge (#1100223026). The XGC1-produced heat-flux has been multiplied by 1.25 in order to match the peak heat-flux value. The experimentally measured footprint has been shifted radially outward by 1.1mm for matching with the XGC1-calculated peak location (in the experimental data, there is always some uncertainty in the magnetic separatrix location). Even though the agreement in the "significant heat-flux region" is reasonably good, some discrepancies in the two tail regions can be noticed: Experimental profiles always have longer heat-flux tails at far scrape region (called "background heat-flux" in Ref. [26]), and often have deeper penetration in the private flux region. The lack of heat-flux tail in the XGC1-produced data at far scrape-off region ($\Delta r \gtrsim 2mm$ in this case) could be from the artificial outer boundary condition used in the XGC1 simulation – the potential is set to zero in the limiter shadows when the magnetic field lines intersect the main chamber limiter wall at other than the divertor plates. Alternatively, it could be from the neglect of the radiative energy deposition in the simulations.

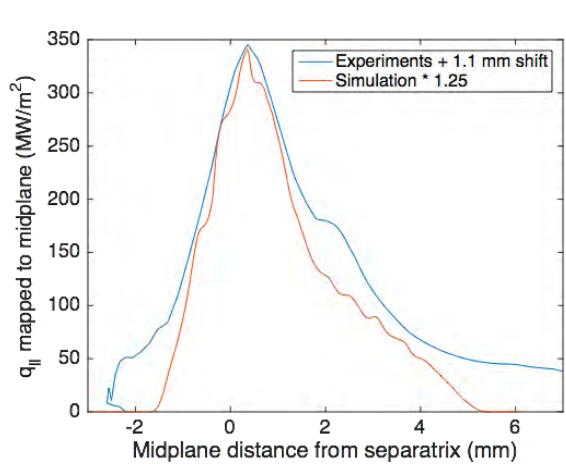

Fig. 9. Comparison of the XGC1-produced divertor parallel heat-flux footprint (red line) against experimental infrared measurement in the C-Mod 0.5MA case (blue line).

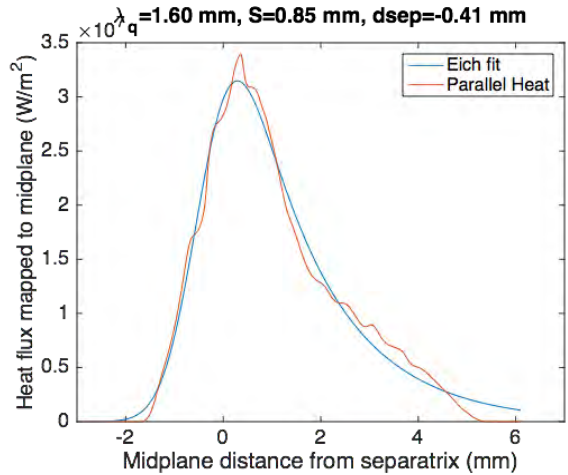

Fig. 10. Eich-formula shape (blue) of the XGC1-calculated (red) parallel divetor heat-flux footprint using Eq. (2), for the Fig. 9 case.

The discrepancy in the private flux region around -2mm is most likely an experimental instrument or analysis issue. In these C-Mod experiments, a highly acute view angle of the IR camera onto the raised divertor plates in the private flux region was likely to give an artificially wider heat-spread detection (C-Mod has improved this issue in their most recent experiments, but those data were not available at the time of this study). In DIII-D, the incidence angle of the magnetic field lines on the private region of the divertor plates is so shallow that the mapping of the parallel heat-flux to the outboard midplane can be almost singular.

Figure 10 is the Eich-formula shape of the XGC1-calculated divertor heat-flux footprint using Eq. (5) for the above C-Mod 0.5MA discharge (#1100223026). The red curve is the XGC1 data and the blue curve is the Eich-formula shape, yielding $\lambda_q$=1.60mm. For comparison, the experimentally obtained $\lambda_q$ is1.55mm and the regression formula (3) yields $\lambda_q^{(14)}$=1.44mm. The Eich-formula shape resembles the XGC1-calcuated shape quite well, not only for this case but also in all other cases.

Similar calculations and analyses have been performed for all the discharge cases listed in Table I. The XGC1-produced $\lambda_q$ values have been added to Fig. 3 of Ref. [18], and shown in Fig. 11 here. Black open inverted-triangle, circles, and squares represent the XGC1-calculated $\lambda_q$ values for NSTX, DIII-D, and C-Mod, respectively. The color marks are the original data points in Fig. 3 of Ref. [18]. All the XGC1-produced $\lambda_q$ values are confined within the experimental regression error bar, thereby validating the gyrokinetic equations, boundary conditions, and the assumption that the electrostatic blobs dominate the turbulence effect in the divertor heat-flux simulation. There remains, though, a possibility for correction to the results by electromagnetic turbulence.

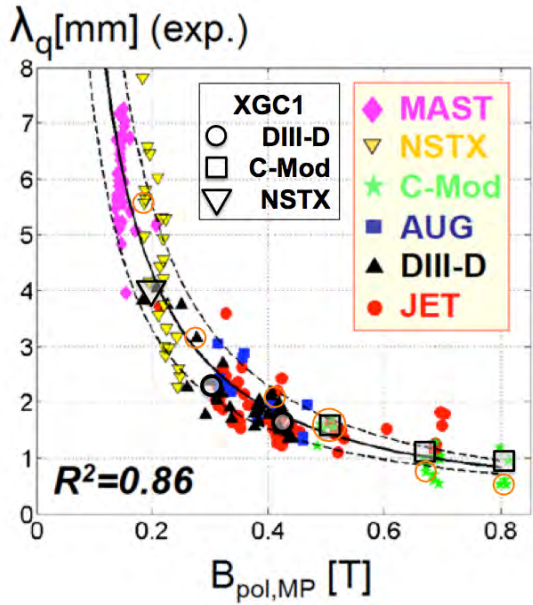

*Fig. 11. XGC1-produced $\lambda_q$ results from all six discharges of Table I have been added to Fig. 3 of Ref. 71. Data points circled in orange represent experimental discharges that are used for simulation.*

In Fig. 11, the data points circled in orange represent experimental discharges that are used for the simulations. The reason why a simulation case has a slightly higher $B_P$ from the corresponding experimental case is that the simulations use the new kinetic-EFIT for magnetic equilibrium reconstruction while the experimental points used the old EFIT [27]. At smaller $B_P$ values in the figure where $\lambda_q$ increases sharply with decreasing $B_P$, this difference in $B_P$ makes a more significant $\lambda_q$ difference between the experimental and simulation points. At the high end of $B_P$ values, where the higher current C-Mod discharges are used, the experimental $\lambda_q$ becomes smaller than the XGC1 results and even smaller than the lower bound of the experimental regression error bar. It can be conjectured that the erratically broadened experimental footprint tail into the private flux region, due to the glancing IR camera view angle as discussed earlier, gives a greater spread factor S in the experimental data and hence reduced experimental $\lambda_q$ in accordance with Eq. (2). Since this IR camera issue has been resolved on C-Mod after the present study was finished, future XGC1 simulations for high current discharges can be highly meaningful, especially at $B_P$ much higher than 0.8T.

The neoclassical type scaling $1/B_p$ of the divertor heat-flux width was reported in Ref. [28] from an earlier XGC simulation result, followed by a later publication [29]. A similar scaling was found later from a heuristic drift-based model [30].

## 4. Prediction for divertor heat-flux width in a model ITER Q=10 plasma

Simulation results for ITER are surprisingly different from those for present experiments. Figure 8 shows the parallel power flux footprint to the outer divertor, mapped to the outboard midplane, in an "XGC1 modeled ITER edge plasma" at $I_p$=15MA with Q=10. From Fig. 8, it can be seen that $\lambda_q$ is 5.9mm with 1.7% standard deviation error, and is ~6 times wider than the prediction from the $1/B_P^{1.19}$ scaling (which yields $\lambda_q$<1mm). Apparently, $\lambda_q$ does not become smaller in proportion to $1/B_P^{1.19}$ in the XGC1-modeled ITER edge plasma, as it does in the present-day tokamaks.

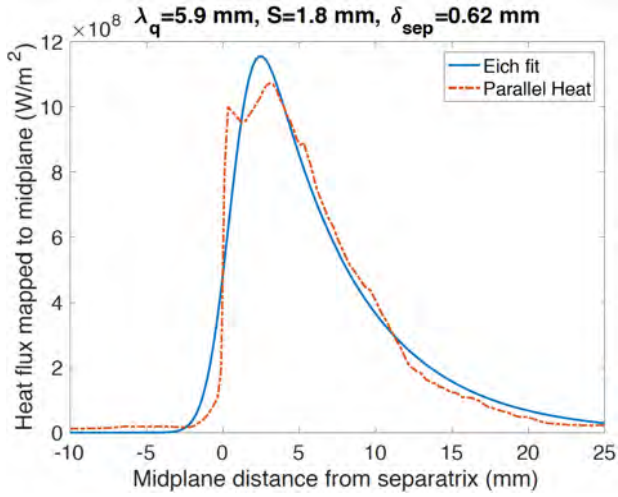

*Fig. 12. Parallel heat-flux footprint on the outer divertor plate, mapped to outboard midplane, in a model ITER plasma edge at 15MA. The "Eich fit" function describes well the XGC1-produced footprint.*

Here, the "XGC1-modeled ITER edge plasma" means that the initial ITER profiles, given to XGC1 from the reduced transport modeling code JINTRAC [31] -- using the assumption that radial transport in-between ELMs is controlled to keep the pedestal close to the linear ideal-MHD stability limits-- was altered in the direction demanded by XGC1 to produce approximate power-balance between the steep edge pedestal and divertor plates with roughly 100MW of power flow to the outer divertor plates, which is similar to the 66 MW expected for Q = 10 operation assuming a 2:1 outer to inner divertor power flow asymmetry. The initial ideal-MHD limited boundary plasma profiles from JINTRAC produced such a high turbulence level in XGC1 around the separatrix that the heat load to the outer divertor plates was about 700MW. The radial energy balance was therefore broken and the pedestal gradients were becoming milder with time in the XGC1 simulation. Therefore, to rapidly reach a plasma equilibrium in the direction demanded by XGC1, which is consistent with the expected power balance between the pedestal and the divertor plates in ITER, we weakened the pedestal density and temperature gradients (and widened the pedestal width at the same time) until about 100MW of power flow across the steep pedestal and a similar power is lost to the outer divertor plates, in such a way that the average plasma pressure in the burning plasma core is similar to the initial pressure (see Fig. 13, showing only the edge region). The adjustment is done only for the plasma profiles, not for the equilibrium magnetic field. Ideally, it would have been better if we let XGC1 find a power balanced edge profiles by consuming a much longer simulation. However, such a computing resources is not available at this time.

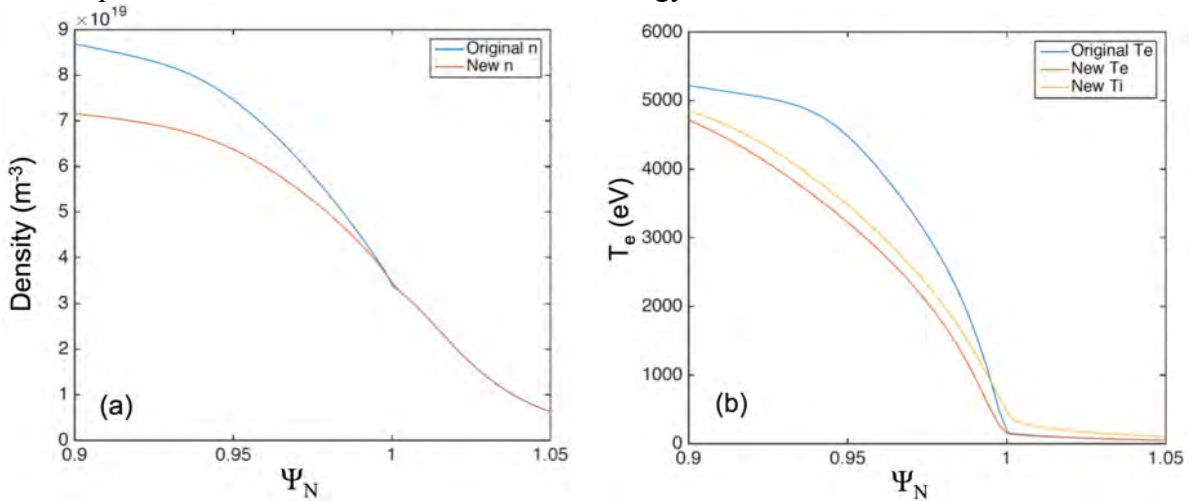

*Fig. 13. Ideal MHD limited edge density and temperature profiles (blue) modelled for an ITER Q =10 plasma by a reduced transport-modeling code JINTRAC and the XGC1 adjusted profiles (orange and yellow) until the total heat flux to the outer divertor plates is ~100MW with an approximate power-balance with the heat flux across the steep pedestal gradient. The ideal MHD limited edge plasma profile yielded too strong an edge turbulence in XGC1 with the total heat flux being too high (~700MW) to satisfy the eventual core-edge radial power balance.*

We remark here that the ITER edge plasma profiles obtained here, even though it is in the direction demanded by XGC1 for gyrokinetic power balance in the edge region of interest, may not be exactly reproducible in the future ITER experiments. In this sense, the ITER study is different from the studies performed on the three US tokamaks in which actual (but rough) experimental profiles have been used as the initial input to the XGC1 simulation. Thus, the prediction made here for ITER needs to be considered with caution.

Figure 14 depicts time behavior of the radial heat flux across the steep pedestal gradient radius ($\Psi_N=0.975$), showing an approximate power balance with the core heating and the divertor power deposition (100MW to outer divertor). Again, we caution that we only have "approximate" power-balance since the simulation has not been performed long enough to achieve the absolute power balance (see Figs. 3 and 4 for comparison and corresponding comments). It can be seen in Fig. 15, as seen in the simulation for the three existing US tokamaks, that the saturation of the heat-flux width is not highly sensitive to the power balance. Figure 15 marks the time window between 0.47ms (vertical solid line) and the end of the simulation time in which the heat-flux footprint in Fig. 12 was measured and averaged.

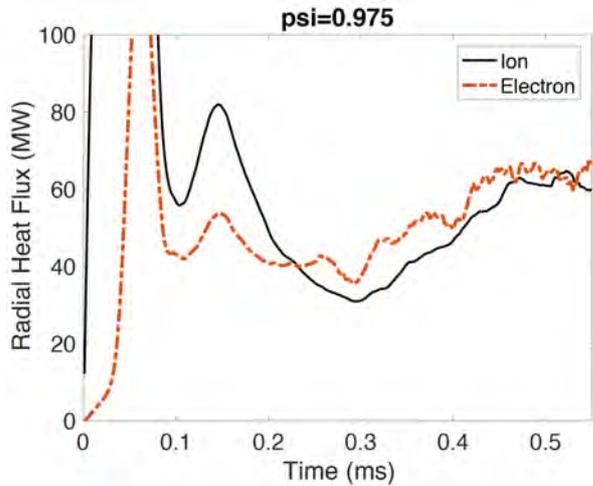 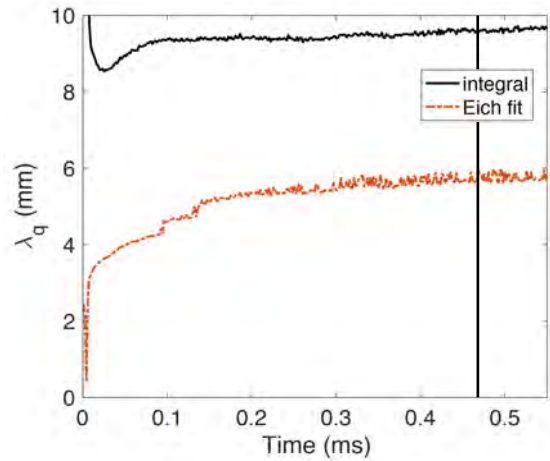

*Fig. 14. Time behavior of radial heat flux across the $\psi_N=0.975$ surface for the model ITER case. Both neoclassical and turbulent heat fluxes are included.*

*Fig. 15. Time behavior of the integral and Eich widths of the divertor heat-flux for the model ITER case. The heat-flux footprint in Fig. 12 is the time averaged shape from t=0.47ms (vertical line) to 0.55ms.*

In Fig. 16, we add the ITER simulation result to Fig. 11 to offer a visual summary of all the XGC1 results from the present study. It may be that ITER's heat-flux width is subject to different physics phenomena from that in present tokamaks. The JET data, marked with a red arrow in Fig. 16, are not studied here and left for future studies.

In search of the radial power balance in the edge region that is consistent with the core heating power of ~100MW, we have manually modified (weakened) the pedestal profile three times in the direction demanded by XGC1. One peculiar feature found during this

**16**

locally power-balanced ITER plasma-profile search process is that even though the turbulence intensity, hence the heat-flux in the steep pedestal, was getting significantly weaker as the edge pedestal gradients were reduced from those MHD limited gradients in the initial JINTRAC results, the divertor heat-flux width $\lambda_q$ was not getting much narrower, as explained earlier. This observation indicates that the divertor heat-flux width may be more closely related to the magnetic field equilibrium geometry than the turbulence intensity, as suggested by Eq. (4) in Sec. 3.B.

The reason why the predicted ITER divertor heat-flux width does not follow the neoclassical ion magnetic-drift trend obtained from the present tokamaks may be found from Fig. 17: The electron contribution completely dominates over the ion contribution both in magnitude and width in ITER. In the present tokamaks, the ion width is always found to be dominant. Even when the electron heat-flux magnitude is larger than that of ions in the high current C-Mod case, the electron width is still narrower than the ion width. In other words, the blobby turbulence effect on the electrons dominates the neoclassical ion orbit effect in both the width and the magnitude in the ITER model plasma, and widens the divertor heat-flux width beyond the very narrow ion magnetic drift width, unlike in the present experiments.

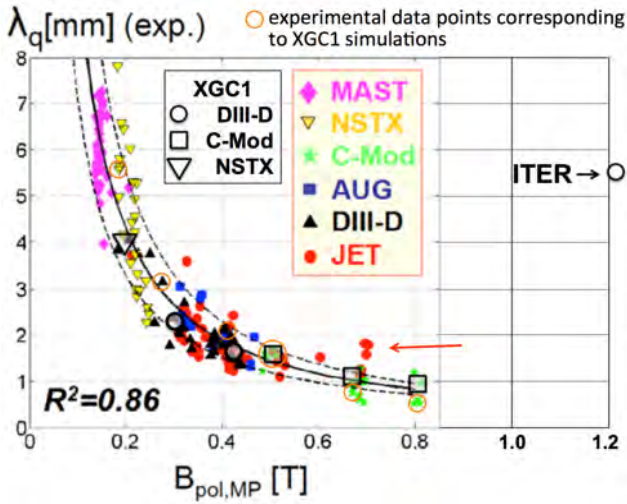
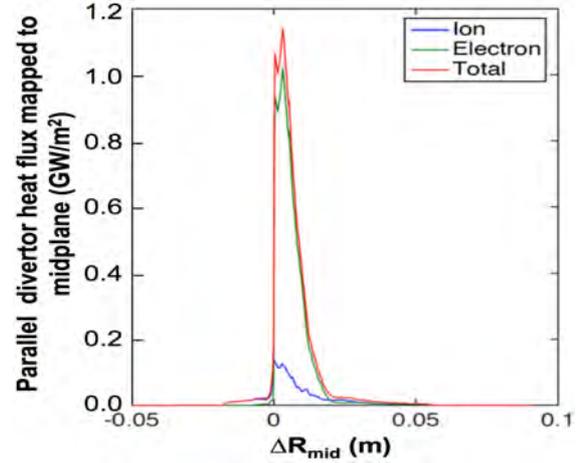

Fig. 16. XGC1 that has reproduced $\lambda_q$ values in the three major US tokamaks predicts $\lambda_q$ =5.6mm in a model ITER plasma edge at $I_P$=15MA.

Fig. 17. Parallel heat flux to ITER outer divertor plates, mapped to outer midplane. $\Delta R_{mid}$ is the midplane radial distance relative to separatrix.

Thus, a possible physics explanation is as follows: Since the size of ITER is much greater than C-Mod, the radial scale length of the magnetic shear is much longer, the sheared ExB flow layer is much wider and, thus, the blob size is much larger than that in C-Mod as shown in Eq. (4). This leads to a much greater turbulence spreading of the electron heat-flux width in ITER than in C-Mod even though the outboard edge ion gyroradius in both devices is similar.

## 5. Conclusion and Discussion

The flux-driven XGC1 edge gyrokinetic code is used for prediction of the heat-flux width on divertor plates in attached divertor plasma conditions. The simulation results are validated (with a possibility for further correction by electromagnetic turbulence effect) against the empirical scaling $\lambda_q \propto 1/B_{pol}^\gamma$ obtained from present tokamak devices, where $\lambda_q$ is the heat-flux width mapped to outbard midplane and $\gamma$=1.19 as obtained by Eich [18, 19], and $B_p$ is the magnitude of the poloidal magnetic field at the outboard midplane separatrix surface.

This empirical scaling predicts $\lambda_q \lesssim 1$mm when extrapolated to ITER, which would require a very high, operationally difficult, separatrix plasma density to achieve semi-detached divertor operation with acceptable divertor power loads in the Q = 10 plasma scenario. XGC1 predicts, however, that $\lambda_q$ for a model ITER plasma edge is ≈5.9mm which enlarges considerably the edge operational space in terms of plasma density and radiative fractions which are compatible with acceptable divertor power loads for the Q = 10 plasma scenario. The physics reason behind this difference is, according to the XGC1 results obtained so far, that while the upper limit of $\lambda_q$ is predominantly set by the neoclassical magnetic drift effect in the present US tokamaks with the turbulence effect playing only a subordinate role to keep the ambipolar plasma flow to the divertor plate through the radial electron spreading, it is predominantly set by the "blobby" nonlinear edge turbulence effect in ITER.

In the high current (0.9MA) C-Mod discharge case, the ion magnetic drift effect again yields a broader footprint than the electron turbulence effect. However, the electron turbulence effect yields a higher level of heat flux. In the ITER case, both the width and the magnitude are dominated by the electron turbulence effect, making the physics distinctively different from the present US tokamaks including the high-current C-Mod case.

We note here again that the divertor heat-flux width $\lambda_q \approx 5.9$mm in the modeled ITER magnetic equilibrium is not highly sensitive to the edge plasma density and temperature profiles, as demonstrated during the plasma profile adjustment process starting from the JINTRAC produced results to achieve a plasma with satisfactory edge power balance in the XCG1 simulations. Only the edge turbulence intensity and the heat-flux magnitude are sensitive to the edge plasma profile. This implies that $\lambda_q$ in ITER could be more sensitive to the radial blob size, than the blob intensity.

The ITER simulation result could suggest that the achievement of cold divertor plasmas and partial detachment with high radiative fractions required for power load and W-impurity source control in ITER high Q scenarios may be more readily achieved and simpler to control than predicted on the basis of the present empirical scaling.

In the future, an extension of the present study to include the electromagnetic turbulence will be performed to quantify possible corrections to the present results, especially for the ITER case. Further identification of the edge turbulence property and the radial transport mechanism will be valuable. Comparison of the gyrokinetic parallel transport mechanisms with the reduced models (e.g., [8, 12-14]) will also be an interesting topic. In particular, a reliable correlation between the midplane plasma profile scale-length and the divertor heat-flux width has not been found from the present study. Study on this issue will be left for a future study. Another interesting subject to be studied includes the role of the quasi-coherent modes, robustly seen in C-Mod discharges, in the physics of $\lambda_q$. For example, why are the quasi-coherent modes not broadening $\lambda_q$ beyond the very narrow width $\lesssim 1$mm in the high current C-Mod discharges?

An important possible bi-product of the present study is that the pedestal density and temperature gradients required to reach the ideal MHD stability criterion at the ITER pedestal may not be reachable due to the strong pedestal turbulence. The gyrokinetically power-balanced ITER pedestal needs to be somewhat wider (see Fig. 13 and the corresponding discussion) and, thus, the pressure gradients lower than those leading to the ideal MHD limit with standard assumptions. This could be another interesting topic to be studied in the future with XGC1 using electromagnetic turbulence. Since XGC1 is a flux-driven full-f (=total-f) code, it can find the power-balanced pedestal plasma profiles in a self-consistent way. There are two interesting possibilities for ITER in this respect: one is that in a micro-turbulence limited ITER pedestal, the ideal MHD unstable pressure limits cannot be reached and thus the edge localized modes (ELMs) may not be of concern; the other one is that the pedestal width may be substantially larger in ITER than in present experiments leading to much higher pedestal pressures being achievable, and thus much higher energy confinement, before the MHD unstable limit is reached and larger-scaled ELMs are triggered.

**Acknowledgement**

We acknowledge valuable discussions with R. Goldston, T. Eich, R. Hager, D. Stotler, M. Churchill, J. Myra, and R. Pitts. This work is funded through the SciDAC program by the U.S. Department of Energy, Office of Fusion Energy Science and Office of Advanced Scientific Computing Research under contract No. DE-AC02-09CH11466 with Princeton University for Princeton Plasma Physics Laboratory. This work is also supported by US DOE through Grant No. DE-FC02-99ER54512 to C-Mod tokamak and the contract No. DE-FC02-04ER54698 to General Atomics. This research used resources of the Oak Ridge Leadership Computing Facility, which is a DOE Office of Science User Facility supported under Contract DE-AC05-00OR22725.

*Disclaimer: ITER is the Nuclear Facility INB no. 174. The views and opinions expressed herein do not necessarily reflect those of the ITER Organization.*

**References**
[1] Kukushkin, A. et al., Jour. Nucl. Mat. **438** (2013) S203
[2] Greenwald, M., Plasma Phys. Control. Fusion 44 (2002) R27
[3] Chang, C.S., Ku, S., Diamond, P.H., Lin, Z., Parker, S., Hahm, T.S., and Samatova, N., Phys. Plasmas **16** (2009) 056108
[4] Ku, S., Chang, C.S., and Diamond, P.H., Nuclear Fusion **49** (2009) 115021
[5] Ku, S., Hager, R., Chang, C.S. et al., J. Comp. Physics 315 (2016) 467-475
[6] C.S. Chang, S. Ku, G.R. Tynan et al., "A fast low-to-high mode bifurcation dynamics in a tokamak edge plasma gyrokinetic simulation, Phys. Rev. Lett, accepted, arXiv:1701.05480
[7] Churchill, R.M., Chang, C.S., Ku, S., "Pedestal and edge turbulence characteristics from an XGC1gyrokinetic simulation", Plasma Phys. Control. Fusion, submitted (2017)
[8] D'Ippolito, D.A., Myra, J.R., and Zweben, S.J., Phys. Plasmas 18 (2011) 060501
[9] Chang, C.S., Ku, S., and Weitzner, H., Phys. Plasmas 9 (2002) 3884
[10] Ku, S.H., and Chang, C.S., Phys. Plasmas **11** (2004) 5626
[11] LaBombard, B., Golfinopoulos, T., Terry, J.L., Brunner, D., Davis, M. Greenwald, M., Hughes, J.W., and Alcator C-Mod Team, Phys. Plasmas **21** (2014) 056108
[12] Stangeby, P.C., "A tutorial on some basic aspects of divertor physics," Plasma Physics and Controlled Fusion **42** (2000) B271
[13] Halpern, F.D., Ricci P., Labit B., I. Furno I., S. Jolliet S. et al., Nucl. Fusion 53 (2013) 122001
[14] Myra J.R. et al and the NSTX Team, Phys. Plasmas 18 (201 1) 012305
[15] Chang, C.S., Ku, S., P. Diamond, P., M. Adams, M., Barreto, R., Chen, et al., Journal of Physics: Conference Series **180** (2009) 012057.
[16] https://www.olcf.ornl.gov/computing-resources/titan-cray-xk7/
[17] DOE FES 2016 Theory/Simulation Target Research, C.S. Chang et al., https://science.energy.gov/~/media/fes/pdf/programdocuments/FES_Theory_Milestone_Report.pdf
[18] Eich, T., et al., Nucl. Fusion 53 (2013) 093031
[19] Eich T. et al 2011 Phys. Rev. Lett. 107, (2011) 215001
[20] Makowski. M.A., Elder, D, Gray, T.K., LaBombard, B., Lasnier, C.J., Leonard, T.W., Maingi, R., T. H. Osborne, Stangeby, P.C., Terry, J.L., and Watkins, J., Phys. Plasmas 19 (2012) 056122
[21] Parker, S.E., Procassini, R.J., Birdsall, C.K., Cohen, B.I., J. Comp. Physics **104** (1993) 41
[22] Diamond, P.H., Itoh, S-I., Itoh, K., Hahm. T.S., Plasma Phys. Control. Fusion **47** (2005) R35
[23] Chang, C.S., Ku, Seunghoe, and Weitzner, H., Phys. Plasmas **11** (2004) 2649
[24] Hahm, T.S., Diamond, P.H., Lin, Z., Itoh, K., Itoh, S-I, Plasmas Phys. Control. Fusion **46** (2004) A323
[25] Wagner F., Nucl. Fusion **25** (1985) 525

[26] Loarte A. et al 1999 J. Nucl. Mater. **266–269** (1999) 587
[27] Lao, L.L., St. John, H., Stambaugh, R.D., Kellman, A.G., and Pfeiffer, W., Nucl. Fusion **25** (1985) 1611
[28] DOE FES Joint Facilities Research Milestone 2010 on Heat-Load Width, Appendix H, http://nstx.pppl.gov/DragNDrop/Program_PAC/JRT_reports/FY2010_JRT_final_report.pdf
[29] Pankin, A., Rafiq, T., Kritz, A.H., Park, G.Y., Ku, S., Chang, C.S. et al., Phys. Plasmas **22** (2015) 092511
[30] Goldston R.J., Nucl. Fusion **52** (2012) 013009
[31] Romanelli, M., Corrigan, Gerard, Parail, V. et al., Plasma and Fusion Research 9 (2014) 3403023